\documentclass[aps,twocolumn,superscriptaddress,amsmath,amssymb]{revtex4-1}
\usepackage{graphicx}  
\usepackage{dcolumn}   
\usepackage{bm}        
\usepackage{verbatim}   
\usepackage{verbatim}   
\usepackage{color} 
\usepackage{ulem}

\begin{document}

\title{Impurity-potential-induced gap at the Dirac point of topological
insulators with in-plane magnetization}
				
\author{M.F. Islam} 
\affiliation{Department of Physics and Electrical Engineering,\\ 
Linnaeus University, 391 82 Kalmar, Sweden} 
\author{Anna Pertsova} 
\affiliation{Nordita, KTH Royal Institute of Technology and Stockholm
University, Sweden} 
\author{C.M. Canali} 
\affiliation{Department of Physics and Electrical Engineering,\\ 
Linnaeus University, 391 82 Kalmar, Sweden}

\date{\today}

\begin{abstract}

The quantum anomalous Hall effect (QAHE), characterized by dissipationless
quantized edge transport, relies crucially on a non-trivial topology of the
electronic bulk bandstructure and a robust ferromagnetic order that breaks
time-reversal symmetry.  Magnetically-doped topological insulators (TIs)
satisfy both these criteria, and are the most promising quantum materials for
realizing the QAHE. Because the spin of the surface electrons aligns along the
direction of magnetic-impurity exchange field, only magnetic TIs with an
out-of-plane magnetization are thought to open a gap at the Dirac point (DP) of
the surface states, resulting in the QAHE.  Using a continuum model supported by
atomistic tight-binding and first-principles calculations of transition-metal
doped Bi$_2$Se$_3$, we show that a surface-impurity potential generates an
additional effective magnetic field which spin-polarizes the surface electrons
along the direction perpendicular to the surface. The predicted gap-opening
mechanism results from the interplay of this additional field and the in-plane
magnetization that shifts the position of the DP away from the $\Gamma$ point.
This effect is similar to the one originating from the hexagonal warping
correction of the bandstructure but is one order of magnitude stronger. Our
calculations show that in a doped TI with in-plane magnetization
the impurity-potential-induced gap at the DP is comparable to the
one opened by an out-of-plane magnetization. 
 
\end{abstract}

\maketitle

\section{Introduction}
\label{into}

TIs are a special class of quantum materials
characterized by a non-trivial bulk band topology that results from band
inversion at time-reversal invariant points of the Brillouin zone (BZ) due to strong
spin-orbit coupling (SOC). TIs are insulating in the bulk but exhibit metallic
behavior at their surfaces \cite{Kane2005,Hasan2010,Bansil2016}. Since the
discovery of TIs, several novel quantum phenomena have been predicted to emerge
when long-range magnetic order is integrated into TIs. Among these are magnetic
monopoles \cite{Qi2009}, giant magneto-optical effects \cite{Tse2010}, and the
dissipationless inverse spin-Galvanic effect \cite{Franz2010}. One of the most
important applications of magnetic TIs is the realization of the QAHE, first predicted 
theoretically \cite{Yu2010} and later confirmed experimentally in Cr- and V-doped 
(Bi$_2$,Sb$_2$)Te$_3$ magnetic TI \cite{Chang2013,Kou2015,Chang2015}. 
Aside from three-dimensional (3D) TIs, the possibility of observing QAHE 
in two-dimensional (2D) buckled honeycomb atomic crystal layers
with an in-plane magnetization has also been explored using microscopic models 
\cite{yafei2016,yafei2017, liu2018}. 

Theoretically, the topological surface states of a 3D TI
near the DP are typically described by a phenomenological
continuum 2D massless Dirac Hamiltonian, which results in a
linear dispersion around the DP.  In this simplified model, time reversal
symmetry breaking terms such as the exchange field in a magnetic TI will
open a gap at the DP only when the direction of magnetization is perpendicular
to the plane of the surface states.  In contrast, an in-plane magnetization
shifts the position of the DP in the direction perpendicular to the
magnetization, without opening any energy gap. On the other hand, for a
realistic system, band structure calculations shows that non-linear corrections
to the Dirac Hamiltonian become important away from the DP. For the TIs of the
Bi$_2$Se$_3$-family, Fu has shown\cite{Fu2009} that the first of these
non-linear corrections is a non-conventional cubic term, known as a hexagonal
warping (HW) term. The HW term allows an in-plane magnetic field or
magnetization to open a topological energy gap at the shifted DP of the surface
states. Based on the Fu model, it has been predicted that QAHE can also be
induced by in-plane magnetic order \cite{Liu2013}.  Theoretical studies of
surface Mn-doped Bi$_2$Te$_3$ \cite{Henk2012topological} and  Bi$_2$Se$_3$
\cite{abdalla2013} TIs based on density functional theory (DFT) indeed confirm
the possible appearance of an energy  gap at the DP for an in-plane
magnetization, in agreement with HW mechanism. More recent first-principles
calculations \cite{Farideh2016} on Sb$_2$Te$_3$ covered with a ferromagnetic Fe
monolayer also find similar small gaps for an in-plane magnetization.  However,
the surface-state energy gap for in-plane magnetization is typically much
smaller ($\approx$ 0.1 meV) than the gap for out-of-plane magnetization
\cite{Henk2012topological}. Based on these results, strong emphasis has been
put on realizing magnetic TIs displaying a magnetic anisotropy that favors an
out-of-plane magnetization \cite{Islam2018}

Besides the coupling with the impurity-induced exchange field, 
the scattering of the surface electrons by the impurity potential can 
also affect the properties of the Dirac surface states by creating impurity resonance states. 
The theoretical approach\cite{biswas2010,balatsky2012} used to address this problem 
is typically based on the usual purely 2D continuum phenomenological model of the surface 
states in the presence of an impurity  potential. The strength of this scattering potential, 
which crucially determines the precise energy position and shape of the resonance states, 
is an unknown parameter. Note that this model is not able to capture the complexity of 
magnetic doping in 3D TIs, whose effect in the density of states strongly depends on the  
TI material and type of impurity. Typically first-principles DFT calculations do not find evidence 
that such resonances occur right at the DP in case of magnetic doping \cite{abdalla2013,canali2014,Farideh2016}. 
Furthermore  experiments\cite{biswas2012,eelbo2014,xu2017,zhong2017,miao2018} seem to indicate 
that these impurity resonances in the density of states reside tens of meV away from from 
the DP. Therefore these impurity resonances are typically not expected to affect the properties 
of the electronic structure right at the DP, although we cannot of course exclude that this might 
happen for some specific situations\cite{sessi2016}.

Some studies based on angular-resolved
photoemission spectroscopy  supported by first-principles calculations, carried
out in non-TI material thin films, such as Ag(111) surface doped with Bi
\cite{Ast2007} or Si(111) surface doped with Ti \cite{Sakamoto2009}, show the
presence of a large spin splitting of the surface electrons in the direction
normal to the surface, which cannot be explained by the ordinary Rashba effect
and is attributed to the impurity potential and SOC.  Essentially, the idea
emerging in these studies is that the Bi-ion doping in  Bi/Ag(111) and the
adsorbed Ti monolayer in  Ti/Si(111) creates a strong {\it in-plane} gradient of
the crystal potential in the surface layer. This gradient, in analogy with the
ordinary Rashba effect, gives rise to an effective magnetic field,
which is directed {\it out-of-plane.} In magnetic TIs, where magnetic order is
induced by doping, the impurity potential may as well result in a non-zero
spin-polarization of the surface electrons normal to the surface. Such a
``magnetic'' influence of the impurity potential on the spin properties of the
Dirac surface states has not been considered before.  As we show below, when an in-plane
magnetization is present, this additional spin polarization can contribute to
opening a gap at the DP in a way similar to the Fu's HW mechanism.
However, in contrast to the very small HW-induced gap, the
impurity-potential-induced gap is of the same size of the gap generated by an
out-of-plane magnetization.  This result implies that the conditions for
realizing the QAHE might be less stringent than what originally anticipated.
Careful manipulation of the doping potential-gradient might also allow a novel
way to enhancing the gap at the DP, leading to the realization of the QAHE at
higher temperatures.

\section{Continuum model}
\label{theory}

The Hamiltonian of the 2D surface states in the
presence of magnetic impurities with arbitrary magnetization, $\textbf{M}$, is
given by
\begin{eqnarray} \begin{split} \mathcal{H_D}(k) =& E_0(k)+
v_f(k_x\sigma_y-k_y\sigma_x)+\lambda(k_+^3 + k_-^3)\sigma_z & \\ & +
\frac{J}{2}{\bf M}\cdot \bm{\sigma} + {\bf B}^i({\bf k})\cdot\bm{\sigma} & \\ =&
E_0(k)-(v_fk_y-\frac{J}{2}M_x -B^i_x({\bf k}))\sigma_x & \\ & +(v_f
k_x+\frac{J}{2}M_y+B^i_y({\bf k}))\sigma_y & \\ & +(\frac{J}{2}M_z +
\lambda(k_+^3 + k_-^3)+B^i_z({\bf k}))\sigma_z \;,& \end{split} \label{2DH}
\end{eqnarray}
where $E_0(k)=\frac{\hbar^2k^2}{2m^*}$. The second term in Eq.~\ref{2DH} is the
usual 2D Dirac Hamiltonian, giving  a linear $\bf k$ dependence around the DP
located at $\bf k= 0$. The third term is the HW term, i.e. a cubic correction to
the linear Dirac Hamiltonian \cite{Fu2009} where $k_\pm=k_x \pm ik_y$. The
fourth term is the exchange coupling between the surface electrons specified by
spin operator $\bm \sigma$ and the impurity magnetic moment $\textbf{M}$, with
${\it J}$ being the exchange energy. The last term corresponds to the exchange
interaction between the surface electrons and an effective magnetic field ${\bf
B}^i$ coming from the potential gradient created by the impurity.  It can be
described phenomenologically as \cite{Sakamoto2009} 
\begin{equation} {\bf B}^i({\bf k})\approx \frac{\hbar^2N}{mc^2\Omega}\int_{\rm
cell}d{\bf r}\; \frac{1}{r} \frac{dV({\bf r})}{dr} u^*_{n{\bf k}}({\bf r}){ \bm
L}\; u_{n{\bf k}}({\bf r})\;, \label{Bnew} \end{equation}
where, $V({\bf r})$ is the electronic potential, $\bm{L}$ is the angular momentum
operator, $u_{n{\bf k}}({\bf r})$ are Bloch functions and $\Omega$ is the volume
of the primitive unit cell containing $N$ atoms. The effective magnetic field ${\bf B}^i$ is proportional to the
average value of the angular momentum ${ \bm L}$ weighted by the potential
gradient. The field ${\bf B}^i$ incorporates the symmetry properties of the lattice via the Bloch 
functions, which is then reflected in the symmetry properties of the band
structure and spin-polarization of the surface states, as clearly evinced in
both experimental studies \cite{Ast2007, Sakamoto2009}. Eq.~\ref{Bnew} also shows that the
${\bf B}^i$ field vanishes at time-reversal invariant ${\bf k}$-vectors, such as
the $\Gamma$ point.

Defining ${\bf M}=(M_{||}\cos\beta, M_{||}\sin\beta, M_z)$, ${\bf
B}^i=(B^i_{||}\cos\gamma, B^i_{||}\sin\gamma, B^i_z)$ and surface wavevector
${\bf k}=(k\cos\theta, k\sin\theta)$, where $M_{||}$, $B_{||}$ are the
projections of ${\bf M}$ and ${\bf B^i}$ parallel to the surface ($x$-$y$
plane), respectively, and the angles are defined relative to the $x$-axis, the
spectrum of the system can be obtained by diagonalizing the Hamiltonian in
Eq.~\ref{2DH}
\begin{eqnarray} 
\begin{split} E({\bf k})= & E_0 \pm \epsilon({\bf k}) & \\
\epsilon({\bf k})=& \sqrt{[v_f^2k^2+B^2_{||}({\bf k})+\frac{J^2}{4}M^2_{||}+J
B^i_{||}({\bf k})M_{||}\cos(\beta-\gamma)} & \\ & \overline{+2v_f B^i_{||}({\bf
k})k \sin(\gamma-\theta) + J v_f M_{||}k \sin(\beta-\theta)} & \\ & \overline{+
(B^i_z({\bf k}) + \frac{J}{2}M_z +\lambda k^3\cos3\theta)^2}]\:. & 
\end{split}
\label{Eigen} 
\end{eqnarray}
It is evident from Eq.~\ref{2DH} that the out-of-plane component of $\bf{M}$
breaks the spin degeneracy at the DP by aligning spins of the surface electron
along the normal, and consequently, a gap opens up at the DP, as evinced by
Eq.~\ref{Eigen}. On the other hand, the in-plane components of magnetization,
$\bf{M}$, only shifts the DP of a TI from the $\Gamma$-point to a point along
the direction perpendicular to M$_{||}$, but does not contribute to the gap
opening at the DP. The amount of the shift of the DP depends on the strength of
the in-plane components of $\bf{M}$. This shift can be also influenced by the
in-plane component of ${\bf B}^i$. 

The HW term in Eq.~\ref{2DH} results in a non-zero out-of-plane component of the
surface spins given by \cite{Fu2009} S$_z({\bf k}) =
\cos(3\theta)/\sqrt{\cos^2(3\theta) + 1/(ka)^4}$, where $a=\sqrt{\lambda/v_f}$.
Note that S$_z(\bf k)$ vanishes at the $\Gamma$ point and its value away from
the $\Gamma$ point is controlled by the strength of the HW term $\lambda$
relative to the Fermi velocity $v_f$. It follows that the combined interplay of
the in-plane magnetization (which shifts the DP away from the $\Gamma$ point)
and the HW-induced S$_z(k)$ can open up a gap at the shifted DP. This was
already pointed out by Fu \cite{Fu2009}. The effective magnetic field
B$^i_z$, caused by the impurity potential, gives rise to a similar mechanism: it
causes an additional out-of-plane spin polarization at $\bf k$ away from the DP.
This impurity-potential-induced spin-polarization can therefore contribute to
the energy gap at a DP which has been shifted by an in-plane polarization. As we
will see below, the effect of B$^i_z$ on the energy gap is typically
larger than the one caused by the HW term.

\section{Results based on atomistic TB method} 
\label{tb}

To support the theoretical
predictions based on the continuum model of Eq.~\ref{2DH}, we used a microscopic
tight-binding (TB) model with parameters fitted to DFT calculations obtained
with Wien2k \cite{Kobayashi2014}.  The TB model for pristine Bi$_2$Se$_3$
includes $s$ and $p$ orbitals and Slater-Koster hopping elements between atoms
in the same atomic layer and between atoms in first and second nearest-neighbor
layers. SOC is incorporated in the intra-atomic matrix elements.  Since the TB
parameters are fitted to DFT calculations, the TB model contains inherently
higher-order corrections to the linear dispersion of the surface states, such as
the HW term.  The TB model is particularly useful for discerning small energy
gaps at DP, which is problematic in DFT. 

For surface calculations with impurity doping, we consider a slab consisting of
six quintuple layers (QLs) and a $3\times{3}$ surface supercell
\cite{Pertsova2014}. An impurity substitutes a Bi atom on the top surface
and is described by a local magnetic moment and an on-site potential
\cite{canali2014}.  The impurity magnetic moment is treated as a classical spin,
modeled by a local exchange field $\textbf{M}$ coupled to the electron spin at
the impurity site. For the non-magnetic part of the impurity potential, we
consider two cases: (\textit{i}) a point-like impurity potential $U$, which acts
as a uniform shift to the on-site energy of the impurity atom; (\textit{ii}) a
realistic impurity potential, which includes the on-site potential $U$ and a
modification of the hopping parameters between the impurity and its neighbors
(in our calculations, all hopping parameters around the impurity are modified by
equal amounts).  The impurity potential $U$ is known to introduce localized 
impurity states that can affect electronic states in the vicinity of the DP
\cite{balatsky2012}.  Here, however, we focus on modifications exactly at the DP. 
For the value of the impurity potential used in this work, the impurity resonant states 
are located in the valence band (approximately 400 meV below the DP of 
the bottom surface states for U=4 eV).
 
The results of TB bandstructure calculations for a particular choice of
parameters are shown in Fig.~\ref{tb} for a point-like (a,b) and a realistic
(c,d) impurity, plus an in-plane magnetic moment. $\mathbf{M}$ is chosen to point in
the $\Gamma$-$K$ direction of the surface BZ. We specifically choose
this direction since the effect of the HW term is identically zero for exchange
fields in the $\Gamma$-$K$ direction \cite{Fu2009}. The bands are plotted in
the perpendicular ($\Gamma$-$M$) direction.  As one can see from
Fig.~\ref{tb}(a), the impurity doping affects the Dirac states of the top
surface (where impurity is located) in several important ways. There is an
expected shift of the top surface DP in momentum space in the direction
perpendicular to $\mathbf{M}$.  In addition, the DP shifts in energy due to the
on-site potential $U$.  Importantly, in the absence of the HW term and for a
point-like impurity, the top surface DP remains gapless [see Fig.~\ref{tb}(b)]. 
Figure~\ref{tb}(b) also shows an 
energy gap of order of $1$~meV resulting from an avoided level crossing between 
the top and bottom Dirac states which are still coupled in a six QLs slab. The
Dirac cone of the bottom surface is unaffected by the impurity.

\begin{figure}[!ht] \includegraphics[width=0.45\textwidth]{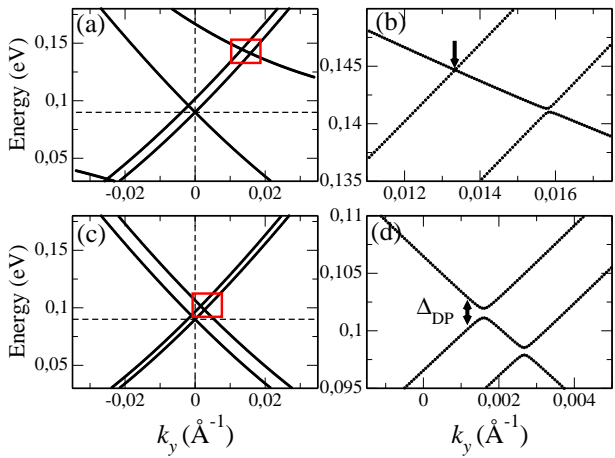}
\caption{TB bandstructure calculations for a six QL Bi$_2$Se$_3$ thin film with
a magnetic impurity on the top surface. (a,b) Point-like impurity with an
in-plane magnetic moment $\bf M$ ($|\bf M| =$ 2 eV) and on-site potential
$U = 4$ eV. (c,d) Realistic impurity with modified hopping parameters between
the impurity site and its neighbor atoms (hopping matrix elements are reduced by
20$\%$). The values of $\bf M$ and $U$ are the same as 
in panels (a,b).  Panels (b) and (d) show small regions of the bandstructures
marked by red squares in panels (a) and (b) respectively.  Dashed lines mark the
position of the bottom surface DP.  Vertical arrow in (b) shows the top-surface
DP. Impurity-potential-induced gap, $\Delta_\mathrm{DP}\approx{1}$~meV,  is
indicated in (d).} \label{tb} \end{figure}

In contrast to the case of a point-like impurity, a more realistic
impurity potential with modified hopping matrix elements will generate an
in-plane electric field giving rise to an effective \textit{out-of-plane}
magnetic field that can gap the DP. This is confirmed by our TB calculations
presented in Fig.~\ref{tb}(c) and (d): the zoom-in of the bandstructure around 
the shifted top surface DP reveals an energy gap,
$\Delta_\mathbf{DP}\approx$1~meV.  Since the HW term is zero for $\mathbf{M}$
pointing along $\Gamma$-$K$, this gap can be attributed solely
to the effective exchange field created by the impurity potential. This is
precisely the effect expected based on continuum-model arguments.  Further
analysis of the bands in 3D momentum space for a few choices of parameters
(hoppings reduced/increased by $20\%$ and $40\%$), confirmed that the DP is
gapped and is located along $\Gamma$-$M$. There is an additional
shift of DP along $\Gamma$-$M$, relative to the shift due to the impurity
magnetic moment.  This suggests that for this specific simplified model of a
realistic impurity potential (and within the momentum resolution of
$0.001\AA^{-1}$), the in-plane projection of the impurity-potential-induced
magnetization is finite and is pointing perpendicular to $\Gamma$-$M$.  

In order to compare the size of the impurity-potential-induced gap with the size of the gap 
due the HW term, we have carried out additional calculations (not shown
here).  For a point-like impurity with magnetic moment along $\Gamma$-$M$, the
resulting gap is solely due to the HW term and is an order of magnitude smaller
than the impurity-potential-induced gap, 
$\Delta_\mathbf{HW}=0.07\approx{0.1}$~meV.  For the case where both the HW term and the impurity-induced magnetic
field are present, namely for a realistic impurity with magnetic moment along 
$\Gamma$-$M$, the gap at DP remains essentially the same as that shown in Fig.~\ref{tb}(d), 
$\Delta_\mathbf{DP}\approx$1~meV. This  
further confirms that the HW gap is much smaller than the
impurity-potential-induced gap.

\section{DFT results} 
\label{dft}

\begin{figure}[!ht] \includegraphics[width=0.49\textwidth]{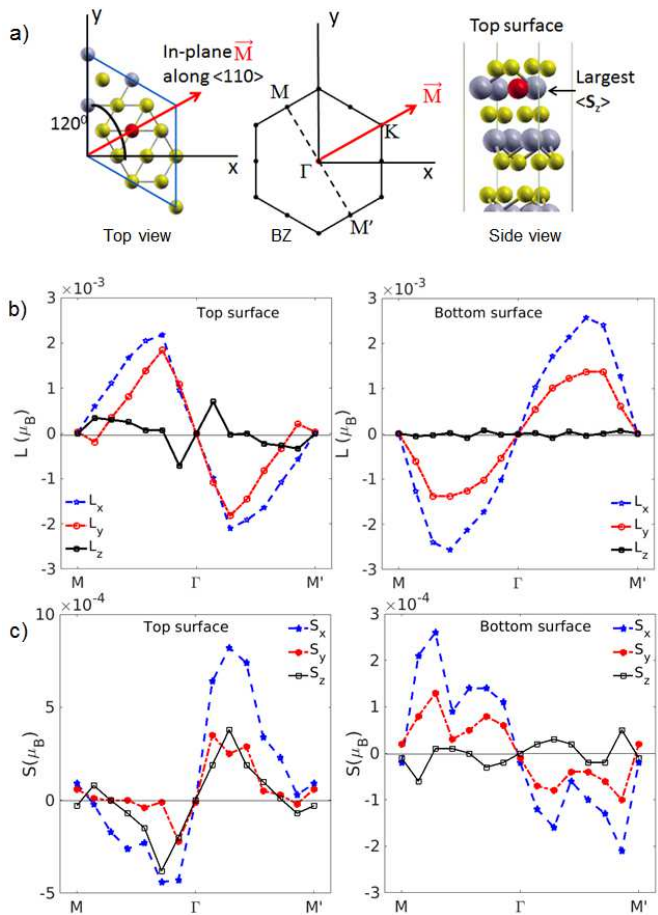}
\caption{
(a) Top and side view of a Bi$_2$Se$_3$ slab, including an impurity atom (red)
on the top surface..
The direction of the in-plane magnetization relative to the coordinate axes is
shown by a red arrow. The BZ shows the M-$\Gamma$-M' (dotted line) path
along which the expectation values of the operators are calculated. (b) and
(c) Expectations values of orbital and spin operators,
respectively, for Fe-doped Bi$_2$Se$_3$. At the top surface, expectation values
are calculated for the Bi atom that is nearest to the Fe impurity. At the bottoms
surface a Se atom is used.} 
\label{ls} 
\end{figure}

To investigate the relative strength of the HW effect and the
impurity-induced field for a realistic system, we have performed 
DFT calculations for Mn- and Fe-doped Bi$_2$Se$_3$ surfaces to 
calculate the surface spin-polarization induced by these two effects. 
The DFT calculations are carried out using the all-electron
LAPW method with PBE exchange correlation functional \cite{Perdew1996} as
implemented in the WIEN2k code\cite{Wien2k}. We have constructed a $2 \times 2 $
surface supercell containing six QLs. The impurity is added by substituting a
Bi atom only from the topmost Bi layer, as shown in Fig.~\ref{ls}a. The doped surface
has been relaxed. To obtain the easy axis, we have calculated the single-ion
anisotropy (SIA) by performing two sets of calculations self-consistently,
including SOC: one with the out-of-plane magnetization and the second one with
the in-plane magnetization along $<$111$>$ ($\Gamma$-K path in the BZ), as
shown in Fig~\ref{ls}. This particular choice of in-plane magnetization is
important for computational purposes, as it preserves one mirror symmetry that
considerably reduces the size of the problem, and allows us to calculate SIA
self-consistently. 

We find that for Mn, the SIA is small (0.08 meV) with an {\it out-of-plane} easy axis.
For Fe, the SIA is considerably larger (1.5 meV) with an {\it in-plane} easy axis.
According to Eq.~\ref{2DH}, the DP remains at the $\Gamma$-point for Mn-doping.
But for Fe-doping, it shifts away from the $\Gamma$-point along M-$\Gamma$-M path
(normal to the magnetization direction), which is confirmed by the TB
calculations discussed above. Since the HW- and impurity-induced field are 
effective only for an in-plane magnetization, here we
consider only Fe-doped Bi$_2$Se$_3$. As mentioned in the discussion of continuum
model, the gap at the shifted DP results from the combined effect of HW and
B$^i_z$. To disentangle these two contributions and study their relative
strengths, we proceed as follows. We note from 
Eq.~\ref{Bnew} that ${\bf B}^i$ has the same dependence as the expectation
value of the orbital momentum operator, $\bf L$. It is convenient to identify
the HW term in Eq.~\ref{2DH} as an internal magnetic field, B$_{HW}(\bf
k)=\lambda(k_+^3 + k_-^3)$, pointing along the z direction. Since the HW term
describes a cubic SOC at the surface of rhombohedral crystal systems, the field
B$_{HW}(\bf k) \sim$ L$_z(\bf k)$. Therefore, the atomic orbital moments, ${\bf L}^{\rm
atom}({\bf k})$, of the surface atoms can provide important insight about the
relative strength of these two contributions and their $\bf k$-dependence. The
HW term affects the Dirac surface states of both the top and the bottom surface of
the slab in the same way. Therefore, we expect that both L$^{\rm atom}_z({\bf
k})$ and S$^{\rm atom}_z({\bf k})$ will be essentially the same for atoms at or
close to the top and the bottom surface. On the other hand, since the  magnetic
impurity is added only at the top surface of the slab, the induced field ${\bf
B}^i_z$ is expected to affect only the Dirac surface states of the top surface,
which will be then reflected in a strong dependence of the local orbital and spin components
of the surface atoms. 

Since Fe doping causes a shift the DP along M-$\Gamma$-M' path, we have
calculated the expectation values of the ${\bf L}$ and the ${\bf S}$ operators
along this path. As discussed above, the impurity-induced field influences only the
atoms at the top surface and is expected to be strong for atoms that are nearest
to the Fe atom. Therefore, we have calculated $<$${\bf L}$$>$ and $<$${\bf S}$$>$ for the Bi
atom located in the same plane of the Fe impurity (see Fig~\ref{ls}a). In order to
distinguish this effect from the HW effect, we have selected a Se atom from the
bottom layer where only the HW effect is expected to play a significant role. The
results are plotted in Fig~\ref{ls}. We note from Fig~\ref{ls}b that
the $k$-dependence of the in-plane components of $<$L$_{x,y}$$>$ is similar and comparable in
magnitude both at the top and the bottom surfaces, except that the sign is
reversed, which is a reflection of the fact that helicity is opposite at the two
surfaces.  On the other hand, it is evident that $<$L$_z$$>$, which is most relevant
for our purpose, is about an order of magnitude larger at $k$-points near $\Gamma$
for the atom at the top surface. Since $<$L$_z$$>$ is a measure of the strength of
the impurity-induced field and the HW effect, we expect its behavior to be reflected on the
spin of the surface electrons, as shown in Fig.~\ref{ls}c.
Evidently, $<$S$_z$$>$, which is responsible for opening a gap at the DP, is an
order of magnitude larger at the top surface. This result clearly indicates that
the impurity-induced field plays a much more dominant role than the
HW term in determining a gap at the DP of the topological surface states for an in-plane
magnetization. 

Based on these DFT results for S$^{\rm atom}_z({\bf k})$, we can try to make an
approximate estimate of the energy gap at the shifted DP induced by the in-plane
impurity potential gradient.  According to Fu's theory the energy gap at the DP
shifted by an in-plane magnetic field is proportional to  S$^{\rm atom}_z({\bf
k}^*)$, where ${\bf k}^*$ is the position of the shifted DP. Previous DFT
calculations \cite{Farideh2016}  find that for an in-plane magnetization, the
gap induced by HW term is $<$ 0.1 meV. Therefore it is not unreasonable to
expect that the in-plane potential gradient caused by impurity can give rise to
a gap $\approx$ ten times larger, that is, of the order a few meV, comparable in
size to the energy gaps obtained in TIs when the magnetization is out of the
plane.     

\section{Conclusions} 

To summarize, based on a continuum model
we have predicted that an in-plane electric potential gradient due to surfaces impurities results in an
out-of-plane effective magnetic field, which can contribute to the
${\bf k}$-dependent spin-polarization of the surface electrons of TIs along with Fu's HW
effect.  Both DFT and TB calculations for Mn- and Fe-doped Bi$_2$Se$_3$ thin
films fully confirmed this prediction and allowed us to elucidate the mechanism
behind this effect.  The most significant result of this analysis is that the
effect of the impurity-potential-induced field is about one order of magnitude larger
than Fu's HW effect (for {\bf k} points close to $\Gamma$-point) and has
significant implications for the gap opening at the DP shifted by the in-plane
magnetization.  Our DFT calculations show that the normal component of the
surface spin-polarization, which is responsible for a gap opening at the DP, is
about one order of magnitude larger for those surface electrons that are
influenced by impurity potential than those that are just affected by the HW
effect.  Thus, for an in-plane magnetization, the impurity-potential-induced
field can open up a gap at the DP of the same order as the one caused by an
out-of-plane magnetization. This important result implies that the
QAHE can be realized efficiently in magnetic TIs irrespective of whether
magnetization is in-plane or out-of-plane to the surface. Thus our work paves
the way for enlarging the domain and enhancing the functionality of magnetic TI
materials suitable for observing the QAHE and other topological spin-dependent
phenomena. One can envisage exploiting this effect to further control and
enhance the magnetic gap at the DP, by engineering doping structures which
generate large in-plane gradient and therefore large out-of-plane induced
spin-polarizations.

{\it Acknowledgments:-} This work was supported by the Swedish Research Council
(VR) through Grant No. 621-2014-4785, and by the Carl Tryggers Stiftelse through
Grant No. CTS 14:178. Computational resources have been provided by the Lunarc
Center for Scientific and Technical Computing at Lund University.

\bibliography{BiSe2019}

\end{document}